\documentclass[pra,amsfonts,amsmath,twocolumn,superscriptaddress]{revtex4}
\usepackage{array,amsmath,amsfonts,amssymb,dsfont}
\usepackage{natbib}
\usepackage[T1]{fontenc} \usepackage{ae} \usepackage{color} 
\usepackage{amsmath}
\usepackage{graphicx}
\usepackage{hyperref}

\begin{document}

\title{Measuring the polarization of electromagnetic fields using Rabi-rate measurements with spatial resolution: experiment and theory}

\author{J.~Koepsell\footnote[1]{These authors contributed equally}\footnote[2]{current address: Max-Planck-Institut fuer Quantenoptik, 85748 Garching, Germany}}
	\affiliation{Department of Physics, ETH Z\"urich, 8093 Z\"urich, Switzerland}
\author{T.~Thiele\footnotemark[1]\footnote[3]{current address: JILA, University of Colorado and NIST, and Department of Physics, University of Colorado, Boulder, Colorado 80309, USA}\footnote[4]{Electronic address: tobias.thiele@colorado.edu}}
	\affiliation{Department of Physics, ETH Z\"urich, 8093 Z\"urich, Switzerland}
\author{J.~Deiglmayr}\affiliation{Laboratorium f\"ur Physikalische Chemie, ETH Z\"urich, 8093 Z\"urich, Switzerland}
\author{A.~Wallraff}\affiliation{Department of Physics, ETH Z\"urich, 8093 Z\"urich, Switzerland}
\author{F.~Merkt}\affiliation{Laboratorium f\"ur Physikalische Chemie, ETH Z\"urich, 8093 Z\"urich, Switzerland}


\renewcommand{\i}{{\mathrm i}} \def\1{\mathchoice{\rm 1\mskip-4.2mu l}{\rm 1\mskip-4.2mu l}{\rm
1\mskip-4.6mu l}{\rm 1\mskip-5.2mu l}} \newcommand{\ket}[1]{|#1\rangle} \newcommand{\bra}[1]{\langle
#1|} \newcommand{\braket}[2]{\langle #1|#2\rangle} \newcommand{\ketbra}[2]{|#1\rangle\langle#2|}
\newcommand{\opelem}[3]{\langle #1|#2|#3\rangle} \newcommand{\projection}[1]{|#1\rangle\langle#1|}
\newcommand{\scalar}[1]{\langle #1|#1\rangle} \newcommand{\op}[1]{\hat{#1}}
\newcommand{\vect}[1]{\boldsymbol{#1}} \newcommand{\id}{\text{id}}

\begin{abstract}
When internal states of atoms are manipulated using coherent optical or radio-frequency (RF) radiation, it is essential to know the polarization of the radiation with respect to the quantization axis of the atom. We first present a measurement of the two-dimensional spatial distribution of the electric-field amplitude of a linearly-polarized pulsed RF electric field at $\sim 25.6\,$GHz and its angle with respect to a static electric field. The measurements exploit coherent population transfer between the $35$s and $35$p Rydberg states of helium atoms in a pulsed supersonic beam. Based on this experimental result, we develop a general framework in the form of a set of equations relating the five independent polarization parameters of a coherently oscillating field in a fixed laboratory frame to Rabi rates of transitions between a ground and three excited states of an atom with arbitrary quantization axis. We then explain how these equations can be used to fully characterize the polarization in a minimum of five Rabi rate measurements by rotation of an external bias-field, or, knowing the polarization of the driving field, to determine the orientation of the static field using two measurements.  The presented technique is not limited to Rydberg atoms and RF fields but can also be applied to characterize optical fields. The technique has potential for sensing the spatiotemporal properties of electromagnetic fields, \textit{e.g.}, in metrology devices or in hybrid experiments involving atoms close to surfaces.
\end{abstract} 		\maketitle

\section{Introduction}
Precise sensing of electromagnetic fields has a vast range of applications, \textit{e.g.}, in establishing SI-traceable standards for the electric field strength~\cite{Sedlacek2012,Holloway2014}, in magnetic field sensing and stabilization in magnetic resonance imaging~\cite{Barmet2008,Duerst2015}, or in the definition of frequency standards by atomic clocks using ultracold atoms in lattices \cite{Ludlow2015,Nicholson2015} or atoms in vapour cells \cite{Micalizio2012}. The quantum nature of atomic systems and their well-understood interaction with electromagnetic fields makes them particularly attractive as sensitive tools for quantum metrology. Quantum-system-based sensors for static or time-dependent magnetic fields have already reached a high degree of maturity and rely on employing systems as diverse as atomic vapors~\cite{budker2007,Affolderbach2015,Horsley2016}, nitrogen-vacancy centers \cite{Thiel2016}, or superconducting quantum-interference devices~\cite{Kleiner2004}. For electric fields, in contrast, the field is still in its infancy. One promising approach is the spectroscopy of atoms in Rydberg states which was shown to be a sensitive tool for the characterization of static and high-frequency electric fields~\cite{Carter2012,Osterwalder1999,Sedlacek2013,Fan2014,Holloway2014,Thiele2014,Thiele2015,Simons2016}.

A wide range of techniques have been developed to measure electromagnetic fields using atomic quantum systems. For example, Ramsey measurements of single trapped ions can detect yN ($10^{-24}$ N) forces originating from very weak electric fields~\cite{Ivanov2016}.
For neutral atoms in Rydberg states, transmission measurements using electrically-induced-transparency have been used to characterize RF electric fields~\cite{Sedlacek2013,Simons2016}. We have recently used pulsed coherent RF Stark spectroscopy to determine the spatiotemporal distribution of static and RF electric fields in a way that is compatible with cryogenic temperatures~\cite{Thiele2014,Thiele2015}.

In this article, we first present our experimental approach to determine, with spatial resolution, the absolute electric field strength of a linearly-polarized RF field $\vec{F}(\vec{r},t)$ and the distribution of the angle $\Theta$ between $\vec{F}$ and a static electric bias-field (Sec.~\ref{sec:experiment}) using an ensemble of Rydberg atoms. Based on these experimental results, we develop a general framework to determine the full polarization ellipse of an arbitrary time-dependent electromagnetic field using the interaction of the field with the internal-state population of atoms (Sec.~\ref{sec:theory}). The framework exploits coherent population transfer in an atomic four-level system in the presence of an adjustable (electric or magnetic) bias-field. We present a set of  equations that can be used to determine the full polarization ellipse of the excitation field from a minimum of five measurements of Rabi frequencies and at least one rotation of the bias-field.

\section{Experiment}\label{sec:experiment}

Our field-measurement technique is based on the observation of coherent population oscillations (Rabi oscillations) between Rydberg states of helium atoms, driven by an \textit{a priori} unknown RF field. The known magnitude of the transition dipole moment $\vec{d}$ allows us to derive an absolute value of the field strength along the direction of the dipole moment from the measured Rabi rate $\Omega$ via the relation
\begin{equation}
\Omega = \frac{\vec{d}\cdot\vec{F_0}}{\hbar},
\label{eq:Omega_from_F_exp}
\end{equation}
with $\vec{F_0}=(F_{x}, F_{y}, F_{z})$ being the vector of amplitudes of the driving field in the laboratory frame. An applied static electric bias-field dominates all other stray static electromagnetic fields and defines the quantization axis of the atom in the laboratory frame. We  use the alignment of the atomic transition dipole with the applied bias-field to obtain information on the polarization vector of the RF field.

The experiments are carried out with the setup described in reference~\cite{Thiele2014} and exploit the imaging techniques developed in reference~\cite{Thiele2015}, as discussed below. A supersonic beam ($v\approx1700\,$m/s) of metastable $(1$s$)^{1}(2$s$)^{1}\,^{1}\text{S}_{0}$ helium atoms propagates in the positive $z$ direction of the laboratory frame $\cal L$ and is excited to the $35\text{s}$ Rydberg state in a field-free region. The Rydberg states are prepared by two successive, resonant one-photon transitions. The first one to the $35\text{p}$ Rydberg state by a $10$-ns-long laser pulse ($\lambda \approx 313$ nm, derived from a frequency doubled dye laser) and the second one by a $140$-ns-long RF pulse of frequency $\omega_{0}/2\pi \approx 25.5655\,$GHz to the $35\text{s}$ Rydberg state. Here, $\omega_{0}$ corresponds to the field-free resonance angular frequency of the $35\text{p}$ to $35\text{s}$ transition, calculated using the energy-dependent quantum defects of helium~\cite{Drake1999}.

\begin{figure}[] \centering \includegraphics[width=86mm]{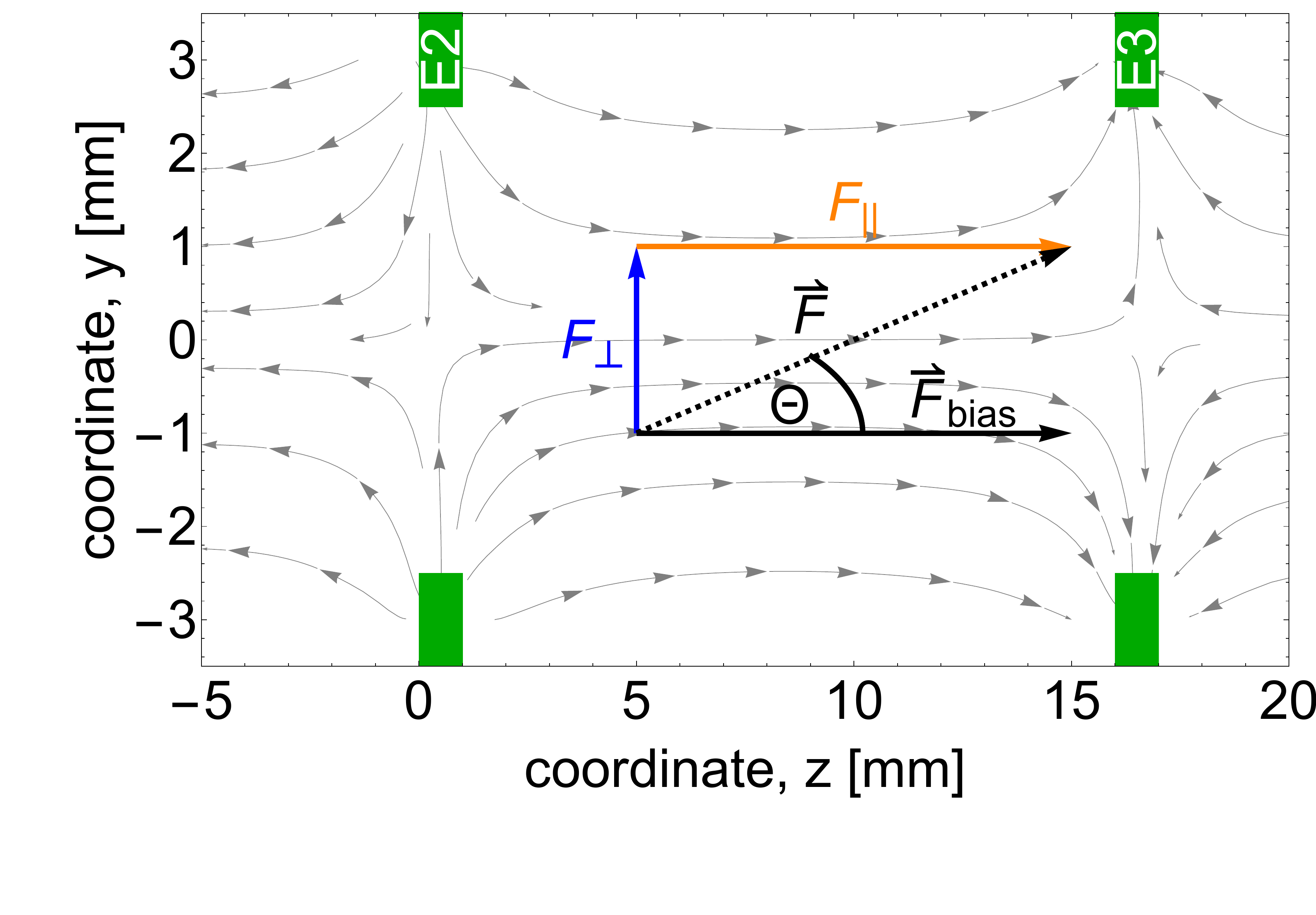}
\caption{Finite-element simulation of the electric field $\vec{F}_{\text{bias}}$ (gray arrows) created by applying a potential difference V$_{23}$ between cylindrical electrodes $E2$ and $E3$ (green boxes). The DC and RF electric-field vectors $\vec{ F}_{\text{bias}}$ and $\vec{F}$ at the position ($y=-1\,$mm, $z=5\,$mm) are indicated by the full and dotted black arrows, respectively. The parallel ($F_{\parallel}$) and perpendicular ($F_{\perp}$) components of $\vec{F}$ with respect to $\vec{F}_{\text{bias}}$ are indicated by the blue and orange arrows.
}
\label{fig:Setup}
\end{figure}

Subsequently, the atoms enter the cryogenic ($T\approx3~\text{K}$) experimental region and propagate through holes in two cylindrical electrodes enclosing our region of interest~\cite{Thiele2014}. A variable potential difference applied to these electrodes results in a static homogeneous electric bias-field $\vec{ F}_{\text{bias}}$ that points in $z$ direction (beam-propagation direction). The bias-field defines the quantization axis for the Rydberg atoms (Fig.~\ref{fig:Setup}) and lifts the degeneracy of the $35\text{p}$ $m=0$ and $|m| = 1$ states.
In the experimental region, the atoms coherently interact with the (to be determined) excitation field, before propagating into the spatially separated detection region. When the atom sample is localized in the most homogeneous part of the electric bias-field  in the middle of the experimental region, the pulsed excitation field $\vec{F}(\vec{r},t)$ transfers population between the $\ket{g_{0}}\equiv \ket{35\text{s},m=0}$ and the $\ket{e_{0}}\equiv \ket{35\text{p},m=0}$ (parallel transition) and $\ket{e_{\pm}}\equiv \ket{35\text{p},m=\pm1}$ (perpendicular transitions) Rydberg states. At this position, $90\%$ of the atom cloud are contained within  $\sim\,1$ mm in the $z$ direction and over $\sim\,$2 mm in the ($x,y$) directions, respectively.

The excitation field is coupled into the experimental region via a horn antenna and has a center frequency $\omega$ with a detuning $\Delta_0=\omega-\omega_0\approx2\pi\times100~\text{MHz}$ from the atomic field-free transition frequency. Its amplitude and phase are controlled using a home-built up-conversion device, operating between $0$ and $50~\text{GHz}$.

The amplitude of the excitation field has a temporal Gaussian envelope of full width half maximum (FWHM) $\Delta t=118\,$ns, truncated to a total pulse duration of $200\,$ns. The FWHM of this pulse in the frequency domain is $7.5$ MHz. The RF-field pulse is short enough such that the atoms only travel $\sim 0.2\,$mm in $z$ direction during the excitation pulse, which is smaller than the longitudinal extent of the atom cloud and small in comparison to the excitation wavelength $\lambda\approx11.7~\text{mm}$. At the same time, the RF-field pulse is long enough to resolve the perpendicular transitions ($\ket{g_0}\leftrightarrow\ket{e_\pm}$) and the parallel transition ($\ket{g_0}\leftrightarrow\ket{e_0}$) spectrally, when an electric DC field $ F_{\text{bias}}$ of $\sim 420\,$mV/cm is applied, causing a Stark-splitting of the transitions by $\sim 30\,$MHz.

The population of atoms in the $35$s Rydberg state is finally detected by pulsed field ionization~\cite{Thiele2014} and the electrons are extracted in the beam propagation direction towards an imaging microchannel-plate-detector assembly. Magnification of the electron pulse by an einzel lens enables precise measurement of the spatial Rydberg-atom distribution in the $xy$ plane. The signal collected in one pixel of the images corresponds to the population of the Rydberg atoms in the $35$s state at a specific location ($x,y$) in the experimental region which can be resolved with $\sim50~\mu\text{m}$ precision~\cite{Thiele2015}.

\subsection{Stark-shifts of $\pi$ and $\sigma$ transitions}\label{sec:Stark_spectroscopy}

\begin{figure}[] \centering \includegraphics[width=86mm]{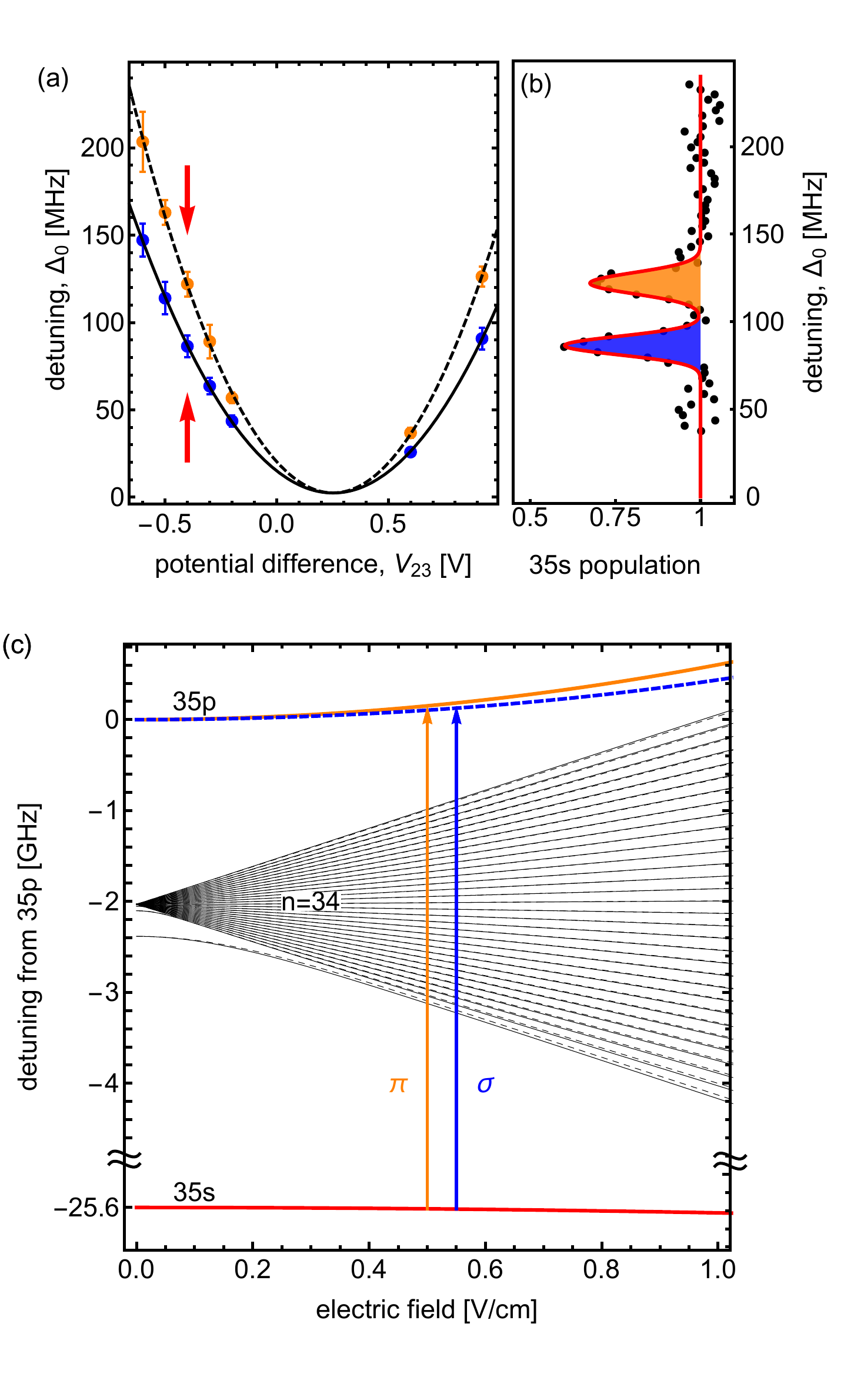} 		\caption{(a) Measured detunings $\Delta_0$ of the center transition frequencies of the parallel (orange datapoints) and perpendicular transitions (blue datapoints) between $35\text{s}$ and $35\text{p}$ as a function of the applied bias potential-difference. The vertical bars indicate the fitted FWHM of the transitions. Black solid and dashed lines are fits to Eq.~(\ref{eq:fit}). (b) RF spectrum of the parallel (orange) and perpendicular (blue) transitions for V$_{23}=-0.4$ V, see red arrow in (a).
(c) Calculated Stark effect of the $(35,s,0)$ (red), $(35,p,0)$ (orange) and $(35,p,\pm1)$ (dashed-blue) states. The energies of the $n=34$, $m=0$ and $\vert m\vert=1$ manifold are indicated by black and dashed-black lines, respectively. The parallel ($\pi$) and perpendicular ($\sigma$) transitions are indicated by orange and blue arrows, respectively.}
\label{fig:identification}
\end{figure}

We have verified that parallel and perpendicular transitions can be independently addressed by the excitation pulse by performing pulsed Stark-spectroscopy. To this end, the bias potential difference V$_{23}$ applied between cylindrical electrodes $E2$ and $E3$ was varied between $-0.6~\text{V}$ and $1~\text{V}$, and the (normalized) total population in $\ket{g_0}$ recorded for detunings $\Delta_0$ between $0~\text{MHz}$ and $200~\text{MHz}$ with respect to the field-free transition [Fig.~\ref{fig:identification}(a,b)].

The two resonances observed in the spectra are well fitted by Gaussian functions, the center frequencies and widths of which are shown in [Fig.~\ref{fig:identification}(a)]. The spectral separation of the resonances increases with increasing $\text{V}_{23}$, \textit{i.e.}, with increasing bias electric-field strength $\vert\vec{ F}_{\text{bias}}\vert$ (called $ F_{\text{bias}}$ hereafter). For $\text{V}_{23}$ between $-0.1~\text{V}$ and $0.5~\text{V}$, the separation of the resonance frequencies is smaller than the Fourier-transform limit of the RF pulses and could not be determined. For larger applied bias-fields, the resonances are broadened by field inhomogeneities.

The observed resonances correspond to the parallel $\ket{g_0}\rightarrow\ket{e_0}$ (orange data points) and to the two degenerate perpendicular $\ket{g_0}\rightarrow\ket{e_\pm}$ transitions (blue data points). On the parallel transition, the population transfer is driven only by the component $F_\parallel$ of the RF field which is parallel to the bias-field (see Fig.~\ref{fig:Setup}). On the perpendicular transitions, the population transfer is driven correspondingly only by the perpendicular component $F_\perp$ of the RF field.  Diagonalization of the single-particle Stark Hamiltonian~\cite{Zimmerman1979} around the $n=34$-manifold states indicates that the frequency shifts of the two transitions scale as $F_{\text{bias}}^2$ for the chosen values of $\text{V}_{23}$ [Fig.~\ref{fig:identification}(c)]. The black and black-dashed lines in Fig.~\ref{fig:identification}(a) are fits to a quadratic function \cite{Osterwalder1999,Thiele2014} for the (field-dependent) detunings $\Delta_0^i( F_{\text{bias}})=\omega_i( F_{\text{bias}})-\omega_0$ ($i=\parallel,\perp$) of the two transitions:
\begin{align}
\label{eq:fit}
\Delta_0^i&=\frac{\alpha_{i}}{2}{ F_{\text{bias},z}}^{2}+\frac{\alpha_{i}}{2}{({F_{\text{bias},x}}^2+{F_{\text{bias},y}}^2)} \nonumber\\
&=\frac{\alpha_{i}}{2}(c*\text{V}_{23}-c*V_{\text{off}})^{2}+\frac{\alpha_{i}}{2}{{F_{\text{bias},xy}}^2}.
\end{align}
Here, $c * \text{V}_{23}$ is the electric field created by the potential difference of V$_{23}$ with a fitted conversion factor $c=0.646(4)\,\text{cm}^{-1}$ averaged over the fit from both transitions. $\alpha_{\parallel}=1330.9~\text{MHz}(\text{V}/\text{cm})^{-2}$ and $\alpha_{\perp}=955.7~\text{MHz}(\text{V}/\text{cm})^{-2}$ are the polarizability differences of both transitions, obtained by numerical calculations~\cite{Zimmerman1979}. $V_{\text{off}}$ and
${F_{\text{bias},xy}}$ account for compensable and noncompensable stray electric fields, respectively. Their values, obtained from an average over the two fits, are $V_{\text{off}}=253(2)~\text{mV}$, and ${F_{\text{bias},xy}}=67(17)~\text{mV}/\text{cm}$. Because the resonance shifts and broadenings induced by $ F_\text{bias}$ are much larger than ac Stark shifts or broadenings of the transitions induced by the RF field, it is justified to assume that the atomic quantization axis is given by the direction of the applied bias-field.

\subsection{Measurement procedure and results}

\begin{figure*}[] \centering \includegraphics[width=\textwidth]{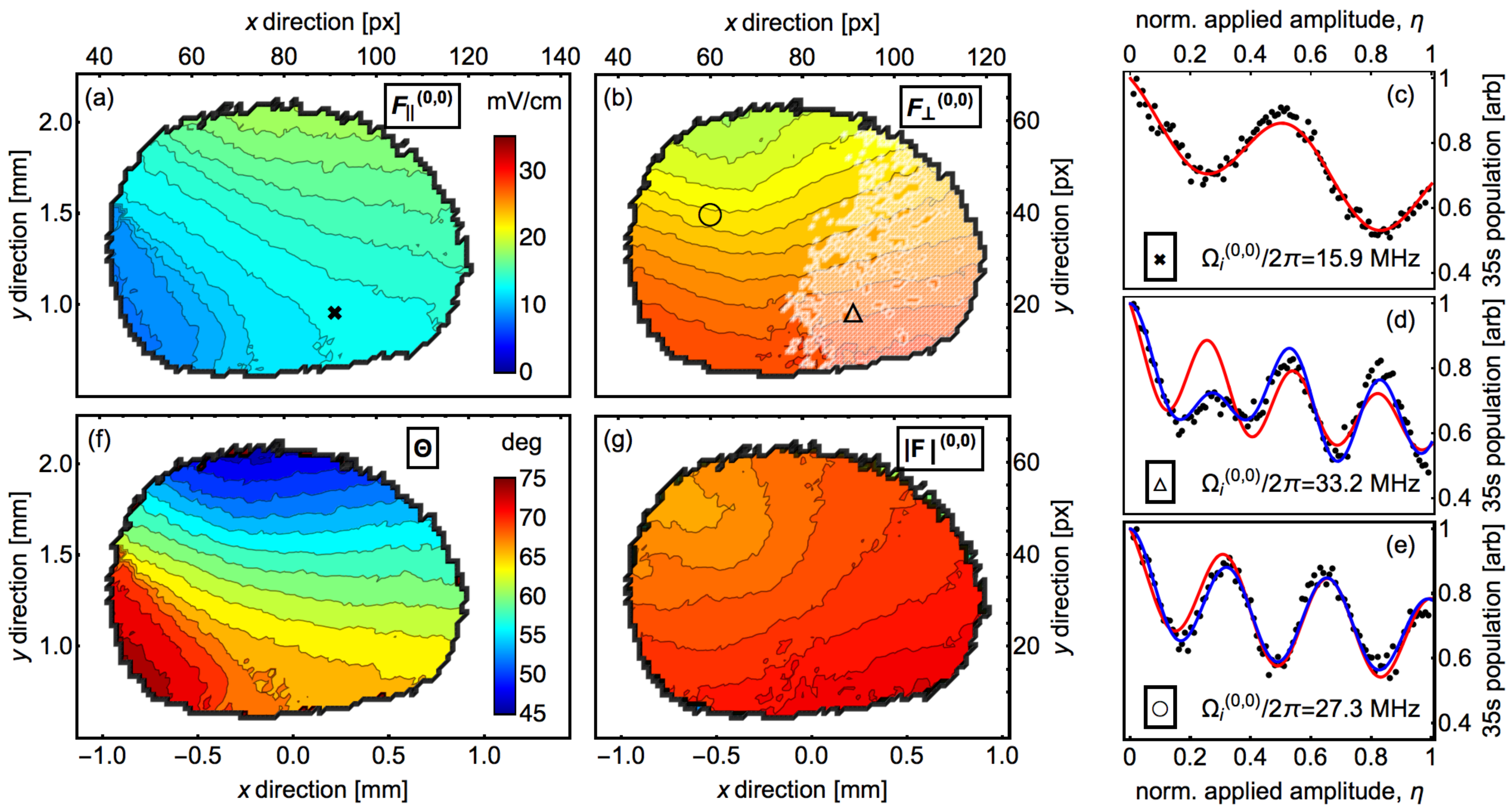}
\caption{Measured field strengths of the RF-field component (a) parallel and (b) orthogonal to the DC electric field for the maximal RF amplitude applied. The separation between adjacent contour lines corresponds to 1 mV/cm. (c) Population in $\ket{g_0}$ for the measurement at the frequency of the parallel transition as a function of the amplitude of the applied pulse at the position indicated by ($\times$) in (a). Data (black dots) and fit of the analytic function [Eq. (\ref{eq:Fitmodel})] (red curve) used to extract $\Omega_{\parallel}$. (d,e) Population in $\ket{g_0}$ for the measurement at the frequency of the perpendicular transition (black dots) as a function of the amplitude of the applied pulse at the positions indicated by ($\triangle$, $\bigcirc$) in (b). (Red curve) fit of [Eq. (\ref{eq:Fitmodel})] to the data and (blue curve) result of an exact numerical simulation (see Appendix~\ref{sec:populationModel}) which indicates a splitting of the perpendicular transition occurring in the white shaded region in (b). The splitting is $3.7\,\text{MHz}$ at the position marked ($\triangle$) in (b). (f) Spatial distribution of the angle $\Theta$ between $\vec{ F}_{\text{bias}}$ and $\vec{ F}$, extracted from measurements presented in (a) and (b). (g) Distribution of total electric field strength $|\vec{ F}|$, inferred from measurements presented in (a) and (b) [color scale as in panels (a) and (b)].}
\label{fig:components}
\end{figure*}

In Fig.~\ref{fig:components}(a,b), we present distributions of the maximum  RF-electric-field strengths $F_\parallel(\vec{r},t)$ and $F_\perp(\vec{r},t)$ in the $xy$-plane in the middle of the experimental region, following the procedure described in reference~\cite{Thiele2015} and summarized in the following. 

We varied the peak amplitude $A$ of the Gaussian-shaped RF excitation pulse from $A_{\text{min}}=0~\text{mV}$ to $A_\text{max}\approx300~\text{mV}$ at the horn antenna, while being resonant with one of the transitions at $\omega_\parallel$ and $\omega_\perp$ for panels (a) and (b), respectively. For each applied RF amplitude $\eta\equiv A/A_{\text{max}}$, the spatial distribution of atoms in $\ket{g_0}$ was recorded and the amplitude-dependent population transfer in each pixel was extracted [Fig.~\ref{fig:components}(c-e)]. We selected the resonances
by changing the frequency of the RF field rather than the bias-field. The frequency change of $\sim30\,\text{MHz}$ is small compared to the central frequency $\omega\approx25.66~\text{GHz}$ of the RF field and does not lead to a measurable change of the RF-field distribution.

By fitting an analytical model (described in Appendix~\ref{sec:populationModel}) to the recorded amplitude-dependent population at every pixel, we extract $\Omega_i(x,y)$ ($i\in\{\parallel,\perp\}$), which is the effective Rabi rate for $\eta=1$ at the maximum of the Gaussian pulse and at the pixel corresponding to the position ($x,y$) in the experimental region \cite{Thiele2015}. From numerical fits to the Rabi oscillations driven on the perpendicular transition [blue data in Fig.~\ref{fig:components}(d,e)], see Appendix~\ref{sec:populationModel}, we learn that the $\ket{g_0}\rightarrow\ket{e_+}$ and $\ket{g_0}\rightarrow\ket{e_-}$-transitions are nondegenerate but split by $\lesssim4~\text{MHz}$ in the light-shaded region because of a small residual  magnetic field on the order of the earth magnetic field. Note, that this could not be resolved by spectroscopic means (measured linewidth of the perpendicular transition $\gtrsim10~\text{MHz}$) but only with the coherent, pulsed spectroscopy we use here. From the good agreement of the numerical fit using identical Rabi rates for the two perpendicular transitions $\Omega_+=\Omega_-$ we conclude that the electric RF field is mostly linearly polarized in the light-shaded region in Fig.~\ref{fig:components}(b). As the remaining data was recorded within a distance $\leq\lambda/10$ we assume that the RF field is also linearly polarized in the other, nonshaded region.

$\Omega_i$ is converted into the electric field strength $ F_i$ using Eq.~(\ref{eq:Omega_from_F_exp}) and the dipole moment $d_\parallel\approx d_\perp=960 \,e a_0$ calculated from the single-particle Stark Hamiltonian~\cite{Zimmerman1979}. The values $\left( F_{\parallel}, F_{\perp}\right)(x,y)$ determined for all pixels (\textit{i.e.}, for all atom positions) represent the two-dimensional distribution of RF electric-field strengths averaged over the extent of the atom cloud in $z$ direction.

We calculate the angle $\Theta$ (see Fig.~\ref{fig:Setup})
\begin{equation} \label{eq:angle}
\Theta=\arctan \left(\frac{ F_{\perp}}{ F_\parallel}\right) \in [ 0,\pi/2 ]
\end{equation}
between the electric RF field and the static electric field [Fig.~\ref{fig:components}(f)] using the data shown in Fig.~\ref{fig:components}(a,b). $\Theta$ varies by more than $27^\circ$ over a distance of $\sim\lambda/5$ ($\lambda\approx11.7~\text{mm}$) in the $xy$-plane, whereas the total electric RF field [Fig.~\ref{fig:components}(g)]
\begin{equation}
\label{eq:Ftotviaperp}
\vert\vec{ F}(t)\vert=\sqrt{ F_{\perp}^2+ F_\parallel^2}
\end{equation}
 varies by $7~\text{mV}/\text{cm}$ ($22\%$ of the maximal value) over the same distance. As detailed in Appendix~\ref{sec:populationModel}, the precision is limited by statistical errors to $\lesssim0.5^\circ$ ($\sim9~\text{mrad}$) and $\lesssim300~\mu\text{V}/\text{cm}$ for the angle and the amplitude measurement, respectively.

The variation of the angle by several degrees and of the RF-field magnitude by $\sim20\%$ over a distance of $\lambda/5$ (\textit{i.e.}, the size of the atom cloud) is compatible with a standing wave between the electrodes separated by $15~\text{mm}$ in $z$ direction, see Fig.~$4$(b) in~\cite{Thiele2014}.
%
%

\section{Theoretical Considerations}
\label{sec:theory}

The experimental approach presented in the previous section facilitates the absolute measurement of the electric-field strength of the RF field with a spatial resolution exceeding by far the wavelength of the radiation. It also allows one to obtain information about the polarization state of the RF field. These conclusions motivated us to investigate theoretically which measurements are required to fully characterize the polarization state of an arbitrary electro-magnetic excitation field, as discussed in the following.

In this section, we first decompose an arbitrary excitation field into linear and circular polarization components in a laboratory frame $\cal L$ and a rotated frame $\cal B$ (Sec.~\ref{sec:excitationfield}). As discussed previously, we relate the amplitudes of these field components to the Rabi rates of transitions in an  atomic four-level system (Sec.~\ref{sec:Rabirates}), consisting of three states which are coupled to a fourth state by $\pi$, $\sigma_{+}$, and $\sigma_{-}$ transitions [see Fig. \ref{fig:PE}(b)]. The labels $\pi$, $\sigma_{+}$, and $\sigma_{-}$ refer to a change of the projection of the atomic angular momentum on the axis of an applied bias-field by 0, $\hbar$, and $-\hbar$, respectively.

In Sec.~\ref{sec:measuring_excitationfield}, we show how the measured Rabi rates can be used to characterize the excitation field using a minimal set of five measurements. Specifically, we discuss the cases in which \textit{i)}  a magnetic bias-field is applied (all transitions can be addressed separately), \textit{ii)} an electric bias-field is applied (only $\pi$- and $\sigma_\pm$-transitions can be addressed separately), and \textit{iii)} no bias-field is applied.

\subsection{The excitation field in $3$ dimensions}
\label{sec:excitationfield}
To characterize a (magnetic or electric) vector field $\vec{ F}(\vec{r},t)$ oscillating at a fixed frequency $\omega$, it can be decomposed into three linear polarization components along $\vec{e}_{j}$ ($j=x,y,z$) in a laboratory-fixed coordinate frame ${\cal L}$, \textit{i.e.},
\begin{equation}
\vec{ F}(\vec{r},t)= \sum_{\substack{j\in \\ \{x,y,z\}}}ˆ{}\frac{1}{2} F_{j}(\vec{r},t)e^{-i(\phi_{j}(\vec{r},t)+\omega t)}\vec{e}_{j}+\text{c.c.} \label{eq:linear}\,\,\,.
\end{equation}
In this case, the excitation field is given by six independent parameters, \textit{i.e.}, the nonnegative amplitudes $ F_{j}(\vec{r},t)$ and phases $\phi_{j}(\vec{r},t)$. The values of these parameters are determined by boundary conditions and are assumed not to change during the measurement time. We can therefore restrict the discussions to a specific point ($\vec{r}=\vec{r}_0, t=t_0$) in space and time and drop the implicit dependence of $F_j$ and $\phi_j$ on $\vec{r}$ and $t$ in the following.

\begin{figure}[]
\centering
\includegraphics[width=86mm]{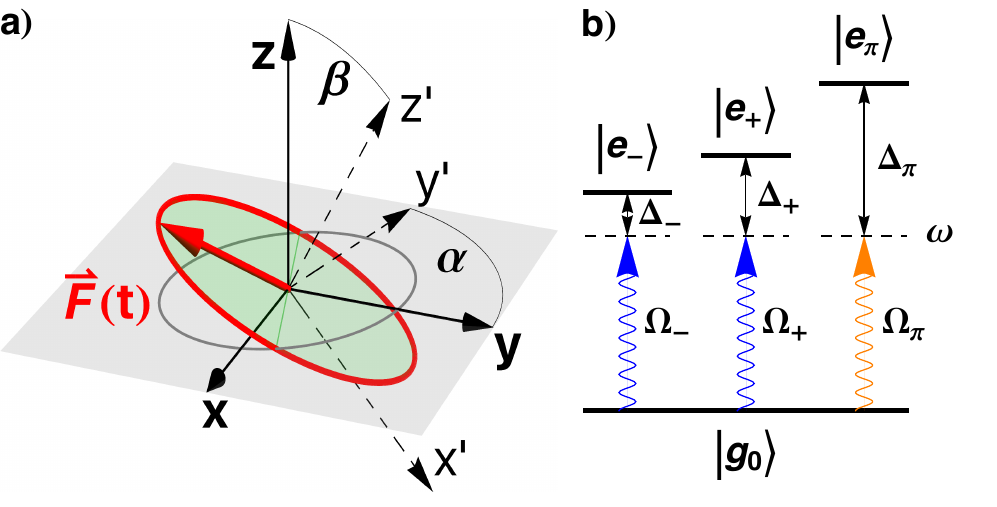}
\caption{a) Polarization ellipse (red) of an excitation field $\vec{ F}(t)$ at $\vec{r}_0$. The angles $\alpha$ and $\beta$ indicate the angles that rotate the laboratory frame $\cal L$ (black) into the bias-field frame $\cal B$ (black, dashed). b) Level diagram of a four-level system with common state $\ket{g_0}$ coupled to three states $\ket{e_{-}}$, $\ket{e_{+}}$, and $\ket{e_\pi}$. The transitions are indicated by blue and orange arrows for $\sigma_{\pm}$ and $\pi$ polarized light, respectively. For each transition [$\gamma= (+,-,\pi)$], the detuning ($\Delta_\gamma$) from the drive frequency $\omega$ and the Rabi frequency ($\Omega_\gamma$) are introduced.}
\label{fig:PE}
\end{figure}

The coherent excitation field $\vec{ F}(t)$ can be represented by a polarization ellipse (PE)~\cite{Saleh2007,Carozzi2000} [see Fig.~\ref{fig:PE}(a)], which is uniquely determined by the three amplitudes $ F_{j}$ and two relative phases~\cite{Alam2015}, the third phase been arbitrarily chosen to be $\phi_z=0$.

The excitation field is conveniently decomposed into a linear $\vec{e}_{\pi}\equiv \vec{e}_{z'}$ and two circular components $\vec{e}_{\pm}\equiv \mp \frac{1}{\sqrt{2}}\left(\vec{e}_{x'}\pm i \vec{e}_{y'}\right)$ of a coordinate frame ${\cal B}=(x',y',z')$ [Fig.~\ref{fig:PE}(a)],
\begin{equation}
\vec{ F}(\vec{r},t)=\sum_{\substack{\gamma\in\\  \{-,+,\pi\}}}ˆ{}\frac{1}{2} F_{\gamma}e^{-i(\phi_{\gamma}+\omega t)}\vec{e}_{\gamma}+\text{c.c.} \label{eq:circular}\,\,\,.
\end{equation}
We always choose the $z'$-direction to be aligned with the quantization axis of the atom, which is conventionally chosen along the direction of an externally applied bias-field. Hence, if the orientation of the bias-field changes, the orientation of $\cal{B}$ with respect to $\cal{L}$ varies. The relative orientation of the two coordinate systems is given by the three Euler angles ($\alpha,\beta,\zeta$) as defined in \cite{Varshalovich1988}. Because the amplitudes $ F_{\gamma}$ that determine the atom-field interaction (see next section) are invariant under rotations around $z'$, we always choose $\zeta=0$ for $\cal{B}$  [Fig.~\ref{fig:PE}(a)] in the remainder of this article.

The field amplitudes in the different coordinate systems can be related through the excitation-field intensity
\begin{equation}
\label{eq_intensity}
\text{I} \propto\sum_{\substack{j\in\\  \{x,y,z\}}}ˆ{} F_{j}^2=\sum_{\substack{j' \in\\  \{x',y',z'\}}}ˆ{} F_{j'}^2=\sum_{\substack{\gamma \in\\  \{+,-,\pi\}}}ˆ{} F_{\gamma}^2,
\end{equation}
and the relations (\ref{eq:amplitude_correspondence}a) and (\ref{eq:amplitude_correspondence}b) resulting from the unitary coordinate transformation
\begin{subequations}
\label{eq:amplitude_correspondence}
\begin{align}
 F_{\pi}^{2}=& F_{z'}^{2} \nonumber\\
=& F_{z}^{2} \cos^2(\beta)+\left( F_{x}^{2} \cos^2(\alpha) +  F_{y}^{2} \sin^2(\alpha)\right)\sin^2(\beta) \nonumber \\
&+ F_{z} F_{x}\sin(2\beta)\cos(\alpha)\cos(\phi_{x})\nonumber \\
&+ F_{z} F_{y}\sin(2\beta)\sin(\alpha)\cos(\phi_{y}) \nonumber \\
&+ F_{y} F_{x}\sin(2\alpha)\sin^2(\beta)\cos(\phi_{x}-\phi_{y}) \\
 F_{\pm}^{2}=&\frac{1}{2}\left( F_{x'}^2+ F_{y'}^2\pm 2 F_{x'}  F_{y'}\sin(\phi_{x'}-\phi_{y'})\right) \nonumber \\
=& \frac{1}{2}\left(\sum_{\substack{j\in\\  \{x,y,z\}}}ˆ{} F_{j}^2- F_{\pi}^{2}(\alpha,\beta)\right) \nonumber \\
&\pm  F_{x} F_{y}\cos(\beta)\sin(\phi_{x}-\phi_{y}) \nonumber \\
&\mp  F_{x} F_{z}\sin(\alpha)\sin(\beta)\sin(\phi_{x}) \nonumber \\
&\pm  F_{y} F_{z}\cos(\alpha)\sin(\beta)\sin(\phi_{y}).
\end{align}
\end{subequations}

\subsection{Interaction with a four-level system}
\label{sec:Rabirates}
The parameters of the excitation field can be determined by measuring the populations at $\vec{r}_0$ in a four-level atomic system consisting of a ground state $\ket{g_{0}}$ coupled to three states $\ket{e_{-}}$, $\ket{e_{+}}$, and $\ket{e_\pi}$ of different frequencies by the radiation at frequency $\omega$ [Fig.~\ref{fig:PE}(b)].

As mentioned before, we choose the quantization axis ($\vec{e}_{z'}$ of $\cal B$) of the atom to be aligned with a constant (magnetic or electric) bias-field $\vec{ F}_\text{bias}$ that dominates the polarization of the atomic transition dipole. The strength of the bias-field can also be used to tune the energy differences between $\ket{g_{0}}$ and $\ket{e_{-}},\ket{e_{+}}$ and $\ket{e_{\pi}}$ and therefore to vary the detunings $\Delta_\gamma\equiv\omega-\omega_{\gamma}$ for $\gamma= (\pi,+,-)$ [Fig. \ref{fig:PE}(b)].

The atom is coupled to the excitation field by the electric- or magnetic-dipole coupling operator ${\cal H}_{\text{int}}=-\hat{\vec{\text{d}}}\cdot\vec{ F}(t)$, $\hat{\vec{\text{d}}}$ being the magnetic- or electric-dipole-moment operator. The amplitudes of the linear and circular components of the excitation field [$ F_{\gamma}$ with $\gamma= (\pi,+,-)$] drive the transitions denoted $\pi$, $\sigma_+$, and $\sigma_-$ in [Fig.~\ref{fig:PE}b], respectively. For each transition, the Rabi rate is given by
\begin{equation}
\Omega_{\gamma}\equiv \frac{\text{d}_{\gamma}~  F_{\gamma}e^{-i\phi_{\gamma}}}{\hbar},
\label{eq:Omega_from_F}
\end{equation}
where $\text{d}_{\gamma}=\bra{e_{\gamma}}\hat{\text{d}}\cdot \vec{e}_{\gamma}\ket{g_{0}}$ is the transition dipole matrix element.

In a frame rotating around $z'$ with angular frequency $\omega$ and using the rotating-wave approximation, the Hamilton-operator of the coupled system in the basis $\left(\ket{g_{0}},\ket{e_{-}},\ket{e_{+}},\ket{e_{\pi}}\right)$ is given by
\begin{equation}
{\cal H}=\frac{\hbar}{2}
\begin{bmatrix}
0& \Omega_{-}^{*}& \Omega_{+}^{*} & \Omega_{\pi}^{*} \\
\Omega_{-}& 2\Delta_{-}& 0& 0\\
\Omega_{+}& 0&2 \Delta_{+}& 0\\
\Omega_{\pi}& 0& 0& 2\Delta_{\pi}\\
\end{bmatrix},
\label{eq:Hint}
\end{equation}
where the detunings $\Delta_\gamma$ are assumed to be much smaller than $\omega$, which is valid for a near-resonant excitation field.

From now on and for the sake of simplicity, we restrict the discussion to situations in which different sets $S$ of transitions are resonant with the excitation field $S\equiv\{\gamma:\Delta_{\gamma}=0 \}$ for a given orientation of $\cal{B}$ w.r.t $\cal{L}$. The other transitions are far detuned ($\Delta_{\gamma^{'}} \gg |\Omega_{\gamma^{'}}|$ for $\gamma^{'} \not\in S$). For a given $S$, the solution to the time-dependent Schr\"odinger equation for an atom initially in $\ket{g_0}$ and described by Eq.~(\ref{eq:Hint}) reveals oscillations of the population in $\ket{g_0}$ with an effective frequency
\begin{equation}
\label{eq:general_Rabirate}
\Omega_{g_0}^{\text{eff}}=\sqrt{\sum_{\gamma \in S}\vert\Omega_\gamma\vert^2}=\frac{1}{\hbar}\sqrt{\sum_{\gamma \in S} \left\vert\text{d}_{\gamma}~ F_{\gamma} \right\vert^2} \qquad
\end{equation}
independent of the phases $\phi_{\gamma}$. The effective Rabi rate ($\Omega_{g_0}^{\text{eff}}$) can be obtained, \textit{e.g.}, by observing the time-dependent population in the state $\ket{g_0}$ (Rabi-oscillations) or from an analysis of measured line shapes (\textit{e.g.}, in case of an observable Autler-Townes splitting~\cite{Holloway2014}). As will be discussed in the next section (Sec.~\ref{sec:measuring_excitationfield}), the nature of the bias-field determines how many different $S$ are available for a given direction of $\cal{B}$, and the minimum set of rotations of the bias-field required to determine the excitation field.

\subsection{Measuring the excitation or bias-field}
\label{sec:measuring_excitationfield}
In this section, we describe how a number of five measurements for $2\leq k\leq 5$ orientations of the quantization axis (\textit{i.e.}, orientations of $\cal B$) are sufficient to determine all excitation field parameters $ F_x, F_y, F_z,\phi_x,\phi_y$ in the laboratory frame $\cal L$.

In Eqs.~(\ref{eq:amplitude_correspondence}a) and (\ref{eq:amplitude_correspondence}b), the field amplitudes $ F_\pi$ and $ F_\pm$ in $\cal B$ are expressed as a function of all five field parameters $ F_x, F_y, F_z,\phi_x,\phi_y$ in $\cal L$. Using Eqs.~(\ref{eq:Omega_from_F}) and (\ref{eq:general_Rabirate}), $ F_\pi$ and $ F_\pm$ can be determined from the measured oscillation frequency $\Omega_{g_0}^{\text{eff}}$ of the population in $\ket{g_0}$. Therefore, it suffices to measure $\Omega_{g_0}^{\text{eff}}$ for $k$ orientations of the bias-field, such that we obtain $5$ (independent) equations from Eqs.~(\ref{eq:amplitude_correspondence}a) and (\ref{eq:amplitude_correspondence}b). These equations can then be solved to find the excitation-field parameters.

The number of bias-field orientations $k$ required to obtain the full information about the excitation field depends on the maximal number of different sets $S$ that are available for the applied bias-field, \textit{i.e.}, the maximal number of discriminable (nondegenerate and resolvable) transitions which is given by the nature of the applied bias-field. In addition, the explicit choice of orientations has to take into account the periodic character of the trigonometric functions in Eqs.~\ref{eq:amplitude_correspondence}(a) and \ref{eq:amplitude_correspondence}(b). Otherwise an underdetermined system of equations might result. Furthermore, the intensity of the excitation field $\vec{ F}$ [Eq.~\ref{eq_intensity}] is independent of the bias-field orientation. As soon as the intensity is determined, the maximal number of independent measurements in a fixed direction of $\cal B$ is  reduced by one.

When a magnetic bias-field is applied to the atom, all transitions ($\pi$, $\sigma_{+}$, $\sigma_{-}$) can be resolved and separately tuned into resonance ($\Delta_\gamma=0$) by the bias-field, \textit{i.e.}, we have a maximum of three sets $S$, each containing one transition. From  a measurement of $\Omega_{g_0}^{\text{eff}}$ for a specific $S$, we thus directly infer $ F_{\gamma}$ for $\gamma=(\pi,+,-)$ [Eq~(\ref{eq:general_Rabirate})]. In order to determine $\vec{ F}(t)$ with a magnetic field applied, it is therefore sufficient to measure $5$ independent values $ F_{\gamma}^{(\alpha,\beta)}$ in only two directions ($\alpha_{i},\beta_{i}$), $i =1,2$ of the bias-field ($\cal{B}$). For instance, from the measured field amplitudes $\left( F_{\pi}^{(0,0)}, F_{+}^{(0,0)}, F_{-}^{(0,0)}, F_{+}^{(0,\pi/2)}, F_{-}^{(0,\pi/2)}\right)$ it is possible to reconstruct $\vec{ F}(t)$ using
\begin{subequations}
\begin{align}
 F_{x}=\sqrt{{\left[ F_{+}^{(0,\pi/2)}\right]}^{2}+{\left[ F_{-}^{(0,\pi/2)}\right]}^{2}-{\left[ F_{\pi}^{(0,0)}\right]}^{2}} \\
 F_{y}=\sqrt{{\left[ F_{+}^{(0,0)}\right]}^{2}+{\left[ F_{-}^{(0,0)}\right]}^{2}-F_{x}^{2}} \\
 F_{z}=\left[ F_\pi^{(0,0)}\right] \\
\phi_{x}=\arcsin\left\{\frac{2{\left[ F_{+}^{(0,\pi/2)}\right]}^{2}- F_{x}^{2}- F_{z}^{2}}{2  F_{x} F_{z}}\right\}\\
\phi_{y}=\phi_{x}-\arcsin\left\{\frac{2{\left[ F_{+}^{(0,0)}\right]}^{2}- F_{x}^{2}- F_{y}^{2}}{2  F_{x} F_{y}}\right\}.
\end{align}
\end{subequations}

The number of necessary orientations of the bias-field ($k=2$) is minimal because for both orientations of $\cal B$, we have the maximal number of independent equations to determine the excitation field, as $ F_{\pi}^{(0,\pi/2)}$ can be derived from Eq.~(\ref{eq_intensity}).

When a static electric field is applied in the absence of a magnetic field, the $\sigma_+$ and $\sigma_-$ transitions are always degenerate, \textit{i.e.}, we have two sets of $S$. For $S=\{\pi\}$, we can directly infer $ F_{\pi}^{(\alpha,\beta)}$ from Eq.~(\ref{eq:general_Rabirate}). For $S=\{\sigma_+,\sigma_-\}$ we obtain
\begin{equation}
\label{eq:Fperp}
\Omega_{g_0}^{\text{eff}} = \sqrt{\vert\Omega_-\vert^2+\vert\Omega_+\vert^2}=\frac{d_\pm}{\hbar} \sqrt{ F_{x'}^2+ F_{y'}^2}=\frac{d_\pm}{\hbar} F_\perp,
\end{equation}
 the excitation field amplitude ($ F_\perp$) transverse to the quantization axis, using $d_\pm=d_+=d_-$, Eqs.~(\ref{eq_intensity}) and (\ref{eq:Omega_from_F}). The minimal number of orientations required to obtain $5$ independent equations is in this case four, \textit{e.g.}, $\left( F_{\pi}^{(0,0)}, F_{\perp}^{(0,0)}, F_{\perp}^{(0,\pi/2)}, F_{\perp}^{(0,\pi/4)}, F_{\perp}^{(\pi/2,\pi/4)}\right)$. Two measurements are independent for the first orientation [here: $(0,0)$], and only one for all subsequent orientations of $\cal B$ [Eq.~(\ref{eq_intensity})]. In the experimental part of this article we have used an electric bias-field to gain information about $\vec{F}(t)$ by measuring $F_{\pi}^{(0,0)}$ ($F_{\parallel}$ in Fig.~\ref{fig:components}(a)) and $F_{\perp}^{(0,0)}$ ($F_{\perp}$ in Fig.~\ref{fig:components}(b)).

A situation in which different sets $\cal S$ can be spectrally resolved, while only the Rabi rate of one particular set $\cal S$ can be measured, is inefficient, because five different orientations of $\cal B$ are needed. For example the orientations $\left( F_{\pi}^{(0,0)}, F_{\pi}^{(0,\pi/8)}, F_{\pi}^{(0,\pi/4)}, F_{\pi}^{(\pi/4,\pi/8)}, F_{\pi}^{(\pi/4,\pi/4)}\right)$ are needed when only the $\pi$ transition can be measured. A measurement of the linear polarization component along the three cartesian axes of a fixed coordinate system is therefore not sufficient to determine the full excitation field. 

It is crucial to make sure that the excitation field does not significantly perturb the level structure of the atomic basis states (\textit{e.g.}, by the ac Stark effect). In the extreme case that the atom is fully polarized by $\vec{ F}(t)$, one reencounters the case of no applied bias-field, and we can only retrieve the magnitude $\left|\vec{ F}(t)\right|$ because there is only one dipole strength $d=d_\pi=d_+=d_-$, and the effective Rabi rate is always given by
\begin{align*}
\label{eq:Fperp}
\Omega_{g_0}^{\text{eff}} &=~~\sqrt{\vert\Omega_\pi\vert^2+\vert\Omega_-\vert^2+\vert\Omega_+\vert^2}&\\
&=\frac{d}{\hbar}\sqrt{ F_{\pi}^2+ F_{+}^2+ F_{-}^2}&\\
&=\frac{d}{\hbar}\sqrt{ F_{x'}^2+ F_{y'}^2+ F_{z'}^2}&\\
&=\frac{d}{\hbar}\vert\vec{ F}(t)\vert\propto\sqrt{I}.
\end{align*}

For completeness, we also mention here that it is possible to determine the direction $(\alpha,\beta)$ of an unknown quantization axis, when $\vec{ F}(t)$ is known in $\cal L$: Eqs.~(\ref{eq:amplitude_correspondence}\,a,b) can be used to obtain two independent equations that determine $\alpha$ and $\beta$. This can be achieved either by measuring different transitions (\textit{e.g.}, $\sigma$ and $\pi$), or by varying the polarization of the excitation field. We see possible applications in experiments where atoms in unknown, inhomogeneous stray fields are excited with laser light in the optical frequency domain, \textit{e.g.}, (Rydberg) atoms that are subject to stray electric or magnetic fields emanating from nearby surfaces~\cite{Hogan2012,Hermann-Avigliano2014,Thiele2014,Thiele2015}. Another possible application is the characterization of fictitious magnetic fields appearing in the context of optical micro traps~\cite{yang2008,klink2013,schneeweiss2014}.

\section{Conclusion}

We have measured the two-dimensional distribution of the angle $\Theta$ between a homogeneous static electric field and a linearly-polarized pulsed RF field centered at $\sim 25.66\,$GHz, and with a temporal Gaussian shape of $200\,$ns total pulse duration. To this end, we determined the Rabi rates from direct coherent population transfer between the $(n,l,m)=(35,$s$,0)$ and the $(35,$p$,0)$ or $(35,$p$,\pm1)$ Rydberg states of singlet helium. The use of pulsed coherent population transfer allowed us to further determine a splitting of the $(35,$p$,+1)$ and $(35,$p$,-1)$ transitions resulting from a small residual magnetic field not compensated for in our experimental setup that could not be detected from the excitation spectra. The precision of the excitation-field measurements was better than $1\%$. For the specific RF configuration we examined, the rotation of the RF polarization and the changes of its amplitude are compatible with the standing waves we typically observe in the experimental region in which the atomic state is manipulated by the RF field \cite{Thiele2014}.

In addition, we have presented a formal analysis of this technique and shown that it can be extended to measure the five independent polarization parameters of an arbitrary optical or RF excitation field from a minimum of five Rabi-rate measurements using an atomic four-level system [Fig. \ref{fig:PE}(a)]. Depending on the number of resolved transitions in the four-level system, the quantization axis needs to be rotated at least once, but not more than four times, \textit{e.g.}, by rotating an external bias-field. We have given explicit examples for the most common experimental configurations. 

Combination of this method with the methods presented in reference \cite{Thiele2015} represents a powerful tool for sensing electric-fields. Our setup allows for the determination of the strength and vector distribution of unknown (stray) static electric fields and of the strength and polarization of time-dependent (RF-) electric fields close to surfaces. Field measurements with a dedicated metrology device based on these principles, using an ensemble of (helium) Rydberg atoms, should be fast and simple, as the optimal configurations to minimize the number of measurements
are known and only frequency or amplitude scans are needed. Additionally, the technique is compatible with cryogenic environments and does not alter the properties of close-by surfaces \cite{Thiele2015}. As the transition frequencies of Rydberg atoms are easily tuned with small applied electric fields, there are few fundamental limitations for the parameter range where such measurements can be performed. 

The method we presented in this article could also be applied to other experiments having the ability to determine Rabi rates and to control bias-fields. We have shown coherent population transfer as a sensitive tool to determine the Rabi rates, but a Rabi rate can also be extracted from spectroscopic measurements in principle. The results of this article could be relevant, for example, for determining polarizations or so-called fictitious magnetic fields~\cite{schneeweiss2014} of optical trapping or excitation fields propagating in nano-photonic-crystal structures, which is a prerequisite to trap single atoms within such structures \cite{Goban2012,Goban2014,Thompson2013,Tiecke2014,Goban2015,Hood2016}.

\section{Acknowledgments}
We thank Dr. Ondrej Tkac and Matija Zesko for providing calculations of the energy structure of He Rydberg atoms in combined magnetic and electric fields, and Josef-Anton Agner and Hansj\"urg Schmutz for invaluable help in setting up the experiment.

We acknowledge the European Union H$2020$ FET Proactive project RySQ (grant N. $640378$). Additional support was provided by the Swiss National Science Foundation (SNSF) under project number $20020\_149216$ (FM) and by the National Centre of Competence in Research "Quantum Science and Technology" (NCCR QSIT), a research instrument of the SNSF. 

\appendix

\section{Modeling the population transfer}\label{sec:populationModel}

The amplitude-dependent population transfer [see Fig. \ref{fig:components} (c-e)] is modeled using an analytical approximation found by \citet{Vasilev2004} using a Dykhne-Davis-Pechukas (DDP) approach. For our Gaussian pulse ($\Delta t=118~\text{ns}$), the local effective Rabi rate $\Omega_{\ket{g_0},i}^{\text{eff}}(t)$ ($i=\pi, \pm$) [Eq.~\ref{eq:general_Rabirate}] varies with time and (normalized) applied RF amplitude $\eta$ as
\begin{equation}
\label{eq:fitRabiRate}
\Omega_{\ket{g_0},i}^{\text{eff}}(\eta,t)=\eta~\Omega_i^{(0,0)}~\exp\left(-\frac{t^{2}}{2\Delta t^{2}}\right).
\end{equation}
After the pulse, the population in $\ket{g_{0}}$ is described by the expression $\text{P}_{\text{DDP}}(\Delta_{i},\Omega_{i}^{(0,0)},\eta)$,
as given by Eq.~($59$) [using Eqs.~($44,52$)] in \citet{Vasilev2004} when their Rabi rate $\Omega(t)$ is replaced by $\Omega_{\ket{g_0},i}^{\text{eff}}(t)$.

In extension of this model we fit the population in $\ket{g_0}$, $\text{P}_{\ket{g_{0}},i}$, to the data at each pixel and for both polarizations using the empirical expression
\begin{align}
\nonumber
\text{P}_{\ket{g_{0}},i}=&(1-\text{C})+\text{C}~\exp\left(-\eta~\chi \Omega_i^{(0,0)}/\Omega_{\Gamma}\right)\\
&\times \text{P}_{\text{DDP}}(\Delta_{i},\chi \Omega_i^{(0,0)},\eta)
\label{eq:Fitmodel}
\end{align}
with the fit  parameters $C$, $\Omega_{\Gamma}$, $\Delta_i$ and $\Omega_i^{(0,0)}$. $C$ is a phenomenological parameter, which takes into account that only a fraction of the detected atoms are driven coherently and given that for zero amplitude ($\eta=0$) the detected signal is normalized to one. $\Omega_{\Gamma}$ is an effective parameter for the dephasing, which results from the finite extent of the atom cloud in $z$ direction and the lifetime of the $35$p state. Additionally, we correct $\Omega_i^{(0,0)}$ for the truncation of the applied Gaussian pulse by multiplication with $\chi=1/1.177$, the ratio between the area under the (experimentally applied) truncated Gaussian pulse and the Gaussian pulse assumed in the model [Eq.~(\ref{eq:fitRabiRate})].

For the $\pi$ ($\sigma$) transition, the average over all fitted traces yields $\text{C}=0.50\pm0.12$ ($0.41\pm0.09$). The fitted detunings $\Delta_{\pi}/2\pi= 2.81\pm 0.28\,$MHz ($\Delta_{\pm}/2\pi=1.55\pm0.84\,$MHz) deviate in each pixel only by a maximum of $2\pi\times 1.57\pm0.90\,$MHz ($2\pi\times 1.14\pm0.74\,$MHz) from an independent spectroscopy measurement. This is small compared to the $\sim 10\,$MHz width of the observed lines [Fig.~\ref{fig:identification}(b)]. For the $\pi$ transition, the dephasing was fixed to $\Omega_{\Gamma}/2\pi=28.4\,$MHz, the average value over all pixels obtained from a previous fit where $\Omega_{\Gamma}$ was a variable at every pixel. For the $\sigma$ transition, a stray magnetic field induces spatially varying dephasing as it lifts the degeneracy of the $\sigma_\pm$ transitions (see below). We therefore keep $\Omega_{\Gamma}$ as a free fit parameter in every pixel, resulting in a variation of $\Omega_{\Gamma}$ between $7\,$ and $43$ MHz. However, we observe that the determined Rabi rates $\Omega_i^{(0,0)}$ are independent of $\Omega_{\Gamma}$ to first order.

In the light-shaded region in Fig.~\ref{fig:components}(a), a stray magnetic field with strength on the same order of magnitude as the earth magnetic field lifts the degeneracy between the $\sigma_{+}$ and $\sigma_{-}$ transition by $\lesssim4\,$ MHz. This leads to an interference between the $\ket{g_0}\rightarrow\ket{e_{-}}$ and the $\ket{g_0}\rightarrow\ket{e_{+}}$ transition as confirmed by a fit, relying on a numerical calculation based on Eq.~(\ref{eq:Fitmodel}), to the data at the (representatively chosen) positions indicated by $\triangle$ and $\bigcirc$ in Fig.~\ref{fig:components}(a), respectively [fits are indicated as blue lines in Fig.~\ref{fig:components}(d,e)]. In detail, we replaced $\text{P}_{\text{DDP}}(\Delta_{i},\chi \Omega_i^{(0,0)},\eta)$ in Eq.~(\ref{eq:Fitmodel}) by a numerical simulation of the driven $3$-level system $(\ket{g_0},\ket{e_+},\ket{e_-})$ [Fig.~\ref{fig:PE}(b)] using the correct truncated Gaussian pulse and the same driving strength $\Omega_\perp^{(0,0)}=\sqrt{2}\Omega_+^{(0,0)}=\sqrt{2}\Omega_-^{(0,0)}$ for both transitions. For panel (d) [panel (e)] in Fig.~\ref{fig:components}, $\Omega_\perp^{(0,0)}/2\pi=33.3(1)~\text{MHz}$ [$27.81(6)~\text{MHz}$] deviates by $0.4\%$ [$1.8\%$] only from the value found with the analytical model, \textit{i.e.}, $\Omega_\perp^{(0,0)}/2\pi=33.1(4)~\text{MHz}$ [$27.3(1)~\text{MHz}$].
A deviation of at most $\sim0.5\%$ is attributed to the statistical error of the fitting procedure for $\Omega_i^{(0,0)}$ at every pixel. This is consistent with the (precision-limiting) statistical fluctuations we observe in the final measurements for the angle and the total field. The magnitude of these fluctuations was (over-)estimated by fitting a linear function to the angle (field) data along a $1$-mm-long line in panels (f) [panel (g)], where the measured values increase approximately linearly with the distance. The precision of $\sim0.5^\circ$ ($\approx9~\text{mrad}$) with which we can determine the angle in our measurements and of $\sim300~\mu\text{V}/\text{cm}$ in the electric field measurements are then determined as the maximal absolute difference of the measured data to the linear fit.

\newpage
\newpage
\end{document}